\documentclass[11pt,twoside]{article}
\usepackage{asp2010}

\resetcounters

\begin{document}

\title{Search for Metal Pollution in 81 DA White Dwarfs}
\author{Detlev Koester,$^1$ Boris G\"ansicke,$^2$ Jonathan Girven,$^2$
  and Jay Farihi$^3$
\affil{$^1$ Institut f\"ur Theoretische Physik und Astrophysik,
  Universit\"at Kiel, 24098 Kiel, Germany}
\affil{$^2$ Department of Physics, University of Warwick, Coventry CV4 7AL}
\affil{$^3$ Department of Physics \& Astronomy, University of
  Leicester, Leicester LE1 7RH}}

\begin{abstract}
A total of 82 DA white dwarfs have been observed with the Cosmic
Origins Spectrograph on the Hubble Space Telescope in a snapshot
program.  The targets were selected to be in the
$T_{\rm eff}$ range from 17000 - 25000~K, where optical metal lines
become weak and difficult to detect. Because of the strong Si, C, and
O resonance lines in the UV, this survey has a sensitivity that is
comparable to that of the Keck/VLT searches for CaII~K in cooler white
dwarfs. These objects also have no convection zone and thus very short
diffusion timescales, assuring that accretion is currently ongoing.
The spectra have high resolution and in most cases fairly good
S/N. About 60\% of them show photospheric metal pollution,
predominantly of Si, but in some cases additional metals are
present. We report the results of a preliminary analysis and discuss
the sources of the accreted matter and the possible r\^ole of
radiative levitation.
\end{abstract}

\section{Introduction}
Traces of heavy metals in helium-dominated atmospheres (spectral types
DBZ, DZ) have been known since almost one hundred years -- in fact,
van Maanen 2, one of the three ``classical'' white dwarfs
belongs to this class. Although the majority of WDs has
hydrogen-dominated atmospheres (DA), the story of metal pollution in
these objects is comparatively recent. Consisting for many years only
of the prototype G74-7 \citep{Lacombe.Wesemael.ea83}, the
second and third member of the class were only discovered in 1997
\citep{Koester.Provencal.ea97, Holberg.Barstow.ea97}. Since then the
number has grown very fast to more than 100 today, including the ones
reported here. This difference between helium and hydrogen atmospheres
is purely observational bias: at the same abundance the equivalent
width of the CaII resonance lines in helium is a factor of hundred or
more larger than in a DA, due to the much larger transparency of
helium at temperatures below 20000~K. Large telescopes like the VLT
and Keck were therefore essential for this development.

Metals in these objects cannot be primordial. Gravitational settling
with diffusion timescales of days to millions of years are always
shorter than the evolutionary timescales, and an external source is
required to replenish the heavy elements. Historically the standard
explanation has been accretion from interstellar matter, although
severe problems -- notably the large underabundance of hydrogen in the
accreted matter and the absence of dense clouds in the solar
neighborhood -- were known and discussed since decades.

With the detection of an infrared excess around a significant fraction
of the polluted objects by the Spitzer Space Telescope and other
ground-based IR telescopes the explanation has now completely shifted
towards the accretion from circumstellar dust in a debris disk,
originating from the tidal disruption of rocky material from a
planetary system. This offers a completely new method to study the
elemental composition of extrasolar planetary systems by relating the
photospheric abundances in the white dwarf to the abundances in the
accreted matter using the equations of the accretion/diffusion
scenario \citep[e.g.][]{Koester09}.

For helium-rich and cool hydrogen-rich white dwarfs with extended
outer convection zones the diffusion timescales can be of the order of
a million years and the current status remains uncertain -- beginning
accretion, equilibrium between accretion and diffusion, or accretion
already ended -- with corresponding uncertainties for the composition
of the accreted matter. We (PI Boris G\"ansicke) have therefore
initiated a program to observe a large number of DA white dwarfs with
effective temperatures between 17000 and 25000~K with the COS
spectrograph on the HST (implemented as a snapshot program). These DAs
have no outer convection zones, the diffusion timescales are extremely
short and equilibrium between accretion and diffusion is
guaranteed. The atmospheric abundances together with diffusion
velocities calculated from the atmosphere models directly give the
diffusion flux for each observed heavy element and thus the relative
abundances in the accreted matter.

The wavelength range observed was 1130-1435~\AA, with typical exposure
times of 1000~s. This range includes strong lines of Si and C, and
fairly strong lines of other important elements like O, Al, S, Fe.
Results for four heavily polluted objects have been published in
\cite{Gansicke.Koester.ea12} and details of the observations,
parameter determinations, and synthetic spectra calculations can be
found there. Here we report on 78 additional DAs, but include the
first four in the statistics and figures.

\section{Interstellar absorption}
All objects show interstellar absorption lines of Si and C, and most
additionally from N, S, O, Fe. The most direct indication is the
SiII\,1260 line and the absence or weakness of SiII\,1265. The latter
line has the larger gf value, but its lower level is 0.036~eV above
the real ground level; it is always the stronger line in the stellar
photosphere, but never appears as interstellar line in our
sample. In most cases the photospheric lines (if present) are also
shifted by 0.1-0.3~\AA\ relative to the ISM lines.

The spectral resolution of $\approx 17$~km/s does not allow to
identify individual components in the ISM lines. However, the
equivalent width of the ISM SiII\,1260 line ranges from 21 to
184~m\AA, corresponding to column densities of $10^{12} -
8.8\,10^{12}$ cm$^{-2}$, if interpreted as single line on the linear
part of growth \citep[e.g.][]{Savage.Sembach96}. These column
densities are typical values for the local interstellar cloud (LIC)
surrounding our solar system \citep{Redfield.Linsky04}; very likely
this is the only absorbing ISM material between us and most of our
sample.

\section{Photospheric metal absorption}
49 of 82 or 60\% of our objects show photospheric Si, 18 additionally C,
and 8 further metals. Taken at face value this number is surprisingly
high, as previous estimates \citep[e.g.][]{Zuckerman.Koester.ea03} put
it much lower at 20-25\%, and the clear signature of the connection
with a planetary system, the infrared excess of a debris disk, is only
found in about 1\% of white dwarfs \citep{Debes.Walsh.ea12}. Since the
diffusion timescales are only a few days, the natural assumption would
be that all these objects are currently accreting, with accretion
fluxes for silicon within a factor of 10 of $3\,10^5$ g/s. However,
before accepting such a conclusion, there is a caveat.
\cite{Chayer.Dupuis10} have shown that some elements, in particular Si
and C, can be supported by radiative acceleration in the atmosphere
even at $T_{\rm eff}$ as low as 20000~K. This would mean that
no current accretion is needed, or that the accretion flows can be
much smaller than calculated from the abundances.

\begin{figure}[ht]
\includegraphics[width=0.7\textwidth,angle=-90]{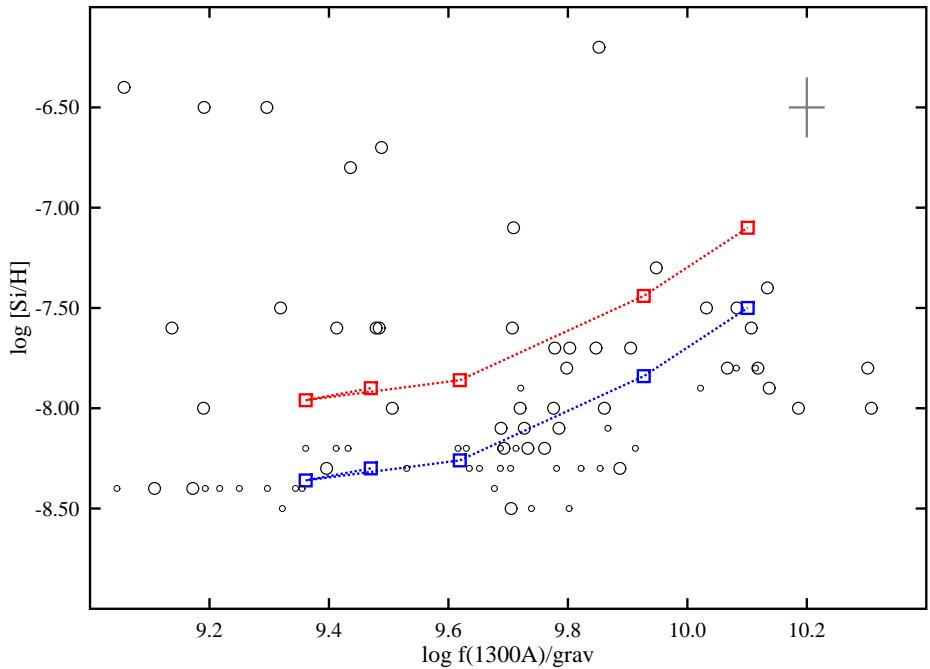}
\caption{log [Si/H] as a function of the parameter, which describes
  the efficiency of radiative levitation (see text). Upper curve with
  (red) squares are predicted abundances for 5 white dwarfs or models,
  lower curve with (blue) squares are the same data shifted by -0.4
  dex. Larger circles are observed Si abundances, the smaller circles
  upper limits. The grey cross in the upper right corner
  indicates typical errors. \label{figsi}}
\end{figure}

Fig.~\ref{figsi} shows the Si abundances as a function of a parameter
calculated from the photon flux ($\log F_\lambda$) minus the surface
gravity $\log g$. We take this as a qualitative measure of the
radiative acceleration on Si, acting against the downward gravitation.
Unfortunately, \cite{Chayer.Dupuis10} publish details only for one
model with $T_{\rm eff} = 20000$~K, $\log g\ = 8$, but predicted
abundances for four real stars can be extracted from a companion paper
\citep{Dupuis.Chayer.ea10}.

All upper limits fall below these prediction. The interpretation in
this scenario is that these stars have never had Si in their outer
layer, or lost it long ago either by downward gravitational settling
or expulsion via a stellar wind.  The other side of the coin is the
conclusion that ALL objects currently showing photospheric Si must
have accreted fairly recently (after levitation became effective
enough to prevent settling) or are accreting right now.  There are a
large number of abundance measurements below the equilibrium line,
which cannot be understood taking the theoretical and observational
data at face value. We note, however, that a better agreement within
the abundance errors for these objects could be achieved by shifting
the prediction downward by about 0.4 dex.  A reason for such an
apparent discrepancy can easily be found.  Fig.~3 in
\cite{Chayer.Dupuis10} predicts very strong abundance gradients within
the atmosphere, whereas our analysis only uses homogeneous
distributions without radiative levitation. A real test of the
levitation predictions could only be done by incorporating the effect
in the model atmospheres and calculate the expected spectra.
Whatever the explanation of such a shift, it would explain the
majority of the Si observations with Si accretion rates $<$ 100~g/s.

15 - 20 objects would remain above the predictions. These could only
be explained by ongoing accretion from circumstellar material and the
accretion rates as determined by the simple description without
levitation would give correct results.

\begin{figure}[ht]
\includegraphics[width=0.7\textwidth,angle=-90]{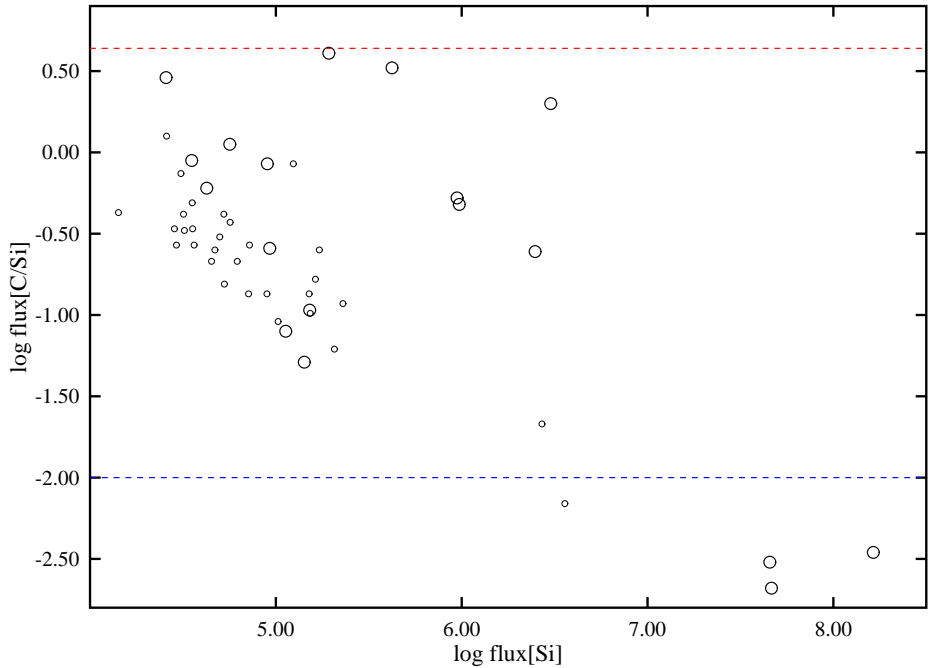}
\caption{Mass flux ratios log [C/Si] as a function of Si mass flux in
  g/s.  Larger circles are observed abundances, smaller circles
  upper limits. The dashed upper (red) line is the
  solar ratio, the lower (blue) line the data for Bulk
  Earth. \label{ratiocsi}}
\end{figure}

\subsection{Carbon to silicon ratio}
There are much fewer objects with C abundance determinations.  The
data in \cite{Dupuis.Chayer.ea10} strongly suggest that carbon should
not be supported in the atmosphere at lower $T_{\rm eff}$. Yet there are
18 cases of observed photospheric C, similar to the number of clear
accretion cases in the discussion of Si. Again, the only viable
explanation for these seems to be ongoing accretion.

Fig.~\ref{ratiocsi} shows the carbon to silicon ratio or upper limits,
as well as those numbers for the sun and Bulk Earth. There
are some values close to the solar/ISM value (possibly accretion from
a low mass companion?) but the majority are a factor of 10 or more
below solar. \cite{Gansicke.Koester.ea12} found that the C/Si ratio of
all four objects was close to the Bulk Earth value (in fact the four
lowest points in the Fig.~\ref{ratiocsi} are from that paper). It
should be noted here that the Si abundance is generally much lower in
the new objects and with a Bulk Earth ratio carbon would not be seen
in these objects (as is shown by the many upper limits). The empty
lower left part of the figure is thus probably observational
bias. That the upper right part is also empty indicates that
very high accretion rates occur only with carbon depleted matter.

\subsection{Source of accreted matter}
The accretion rates in the 15 - 20 objects clearly above the predicted
levitation abundances are also too high to be explained by accretion
from the thin interstellar matter in the local bubble around the
sun. In view of the established connection between infrared excess,
circumstellar dust, and metal pollution in the stellar atmosphere, the
most likely explanation seems to be accretion from a debris disk.

However, that is not necessarily true for the majority of polluted
DAZ. If the levitation calculations are qualitatively correct, the
amount of Si necessary to explain the observations can be extremely
small. The total mass in a 20000~K, $\log g\ = 8$ atmosphere above
$\tau_{Ross} = 1$ is approximately $10^{17}$g. Using data from
Fig.~3 in \cite{Chayer.Dupuis10} the maximum abundance in the outer
layers of such a model is n(Si)/n(H) = $10^{-8}$, and therefore the
mass of Si is smaller than $5.6\,10^{10}$~g.

The cross section for Eddington accretion from interstellar matter,
which is the minimum we would expect, is
\[   a = \frac{2 G M_{wd}}{v^2}\,R_{wd}  = 5.06\, 10^{22}~\mbox{cm}^2 \]
with white dwarf mass and radius $M_{wd}$ and $R_{wd}$, gravitational
constant $G$ and space velocity $v$ (assumed to be 30~km/s). Assuming
the white dwarf crosses just one tiny cloud like our Local
Interstellar Cloud (LIC) with a Si column density of $10^{13}$
cm$^{-2}$, it would sweep up $4.7\, 10^{13}$g of Si. That is orders of
magnitude more than needed to explain the observations, which would in
this scenario diffuse downward, except for the tiny fraction supported
by radiative levitation. 

\section{Conclusions}
All 82 objects show interstellar absorption, very likely due to the
local interstellar cloud (LIC) surrounding the solar system. More
importantly, 60\% also show photospheric Si, and a smaller fraction
also C and a few other metals. If the predictions of radiative
levitation are correct, the majority could be explained by very low
accretion rates that could be possible even within our local
bubble. That leaves about 20\% definitely accreting from circumstellar
material, in agreement with previous estimates. The carbon to silicon
ratio is very low (similar to B.E. values) in the cases of high
accretion rates, and shows a range of values for the lower rates,
although mostly below the solar ratio.


\end{document}